\documentclass[lettersize, journal]{IEEEtran}
\IEEEoverridecommandlockouts
\usepackage{hyperref}
\usepackage{cite}
\usepackage{amsmath,amssymb,amsfonts}
\usepackage{algorithmic}
\usepackage{graphicx}
\usepackage{textcomp}
\usepackage{xcolor}
\usepackage{placeins}
\usepackage{float}
\usepackage{tabularx}
\usepackage{array}
\usepackage{url}
\usepackage{tikz}
\usepackage{pgfplots}
\pgfplotsset{compat=1.18}

\usepackage[caption=false,font=footnotesize]{subfig}
\usetikzlibrary{plotmarks}
\usetikzlibrary{arrows.meta,positioning}

\usepackage{pgfplotstable}

\definecolor{VividGreen}{RGB}{0,175,70}
\definecolor{VividRed}{RGB}{235,35,45}

\pgfplotsset{
resultaxis/.style={
    width=0.82\columnwidth,
    height=0.56\columnwidth,
    xlabel={Beam training overhead (\%)},
    xlabel style={font=\footnotesize},
    ylabel style={font=\footnotesize},
    tick label style={font=\scriptsize},
    xmin=1.6,
    xmax=6.35,
    xtick={2,3,4,5,6},
    grid=major,
    major grid style={
        draw=gray!20,
        line width=0.20pt
    },
    axis line style={
        black!70,
        line width=0.45pt
    },
    tick style={
        black!70,
        line width=0.45pt
    },
    tick align=outside,
    legend style={
        font=\scriptsize,
        draw=black!35,
        line width=0.35pt,
        fill=white,
        fill opacity=0.96,
        text opacity=1,
        inner sep=2pt,
        row sep=0.5pt,
        cells={anchor=west}
    },
    clip=false
},
fixedcurve/.style={
    color=blue!80!black,
    solid,
    line width=0.75pt,
    mark=*,
    mark size=1.75pt,
    error bars/.cd,
    x dir=both,
    x explicit,
    y dir=both,
    y explicit,
    error bar style={line width=0.45pt},
    error mark options={
        line width=0.45pt,
        mark size=1.7pt
    }
},
adaptivecurve/.style={
    color=orange!90!black,
    densely dashed,
    line width=0.75pt,
    mark=square*,
    mark size=1.75pt,
    error bars/.cd,
    x dir=both,
    x explicit,
    y dir=both,
    y explicit,
    error bar style={line width=0.45pt},
    error mark options={
        line width=0.45pt,
        mark size=1.7pt
    }
}
}
\definecolor{fixedpurple}{HTML}{3C08C0}
\definecolor{adaptivegreen}{HTML}{20D500}

\def\BibTeX{{\rm B\kern-.05em{\sc i\kern-.025em b}\kern-.08em
    T\kern-.1667em\lower.7ex\hbox{E}\kern-.125emX}}
\allowdisplaybreaks

\usepgfplotslibrary{statistics}

\definecolor{TauBlue}{RGB}{31,119,180}
\definecolor{TauOrange}{RGB}{255,127,14}
\definecolor{TauGreen}{RGB}{44,160,44}
\definecolor{TauRed}{RGB}{214,39,40}
\definecolor{TauPurple}{RGB}{148,103,189}
\definecolor{TauBrown}{RGB}{140,86,75}
\definecolor{TauPink}{RGB}{227,119,194}
\definecolor{TauGray}{RGB}{127,127,127}
\definecolor{TauOlive}{RGB}{188,189,34}
\definecolor{TauCyan}{RGB}{23,190,207}

\definecolor{FixedBlue}{RGB}{0,78,184}
\definecolor{AdaptiveMagenta}{RGB}{220,0,120}
\definecolor{AdaptiveGreen}{RGB}{0,170,60}
\definecolor{AdaptiveVermilion}{RGB}{230,85,13}

\definecolor{TauBlue}{RGB}{0,114,255}
\definecolor{TauOrange}{RGB}{255,140,0}
\definecolor{TauGreen}{RGB}{0,185,80}
\definecolor{TauRed}{RGB}{255,45,55}
\definecolor{TauPurple}{RGB}{150,55,255}
\definecolor{TauBrown}{RGB}{205,105,0}
\definecolor{TauPink}{RGB}{255,20,147}
\definecolor{TauGray}{RGB}{95,95,105}
\definecolor{TauOlive}{RGB}{170,185,0}
\definecolor{TauCyan}{RGB}{0,195,230}

\definecolor{DistGreen}{RGB}{0,190,60}
\definecolor{DistMagenta}{RGB}{210,0,210}
\definecolor{DistRed}{RGB}{255,40,40}
\definecolor{DistBlue}{RGB}{0,110,255}

\begin{document}

\title{Confidence-Guided Adaptive Beam Selection for Vision-Aided Beam Tracking in mmWave Systems
}

\author{Sohani Munteha Hiam, Mohammad Soleymani, \emph{Senior Member, IEEE}, \\and Rafael F. Schaefer, \emph{Senior Member, IEEE}
 \\ 
\thanks{This work was supported in part by the German Federal Ministry of Research, Technology and Space (BMFTR) through the Transfer Hub \emph{6G-life} under Grant 16KIS2413K and in part
by the German Research Foundation (DFG) as part of Germany's Excellence Strategy EXC 2050/2 -- Project ID 390696704 -- Cluster of Excellence \emph{``Centre for Tactile Internet with Human-in-the-Loop'' (CeTI)}.}
\thanks{ 
Sohani Munteha Hiam is with Universit\"at Paderborn, Germany (email: \protect\url{sohani@mail.uni-paderborn.de}, \protect\url{mohammad.soleymani@uni-paderborn.de}).}
\thanks{ 
Mohammad Soleymani is with Universit\"at Paderborn, Germany, and will join Brunel University of London, UK (email: \protect\url{mohammad.soleymani@uni-paderborn.de}).}
\thanks{
Rafael Schaefer is with the Chair of Information Theory and Machine Learning, Technische Universität Dresden, Germany (e-mail: \protect\url{rafael.schaefer@tu-dresden.de}).
}}

\maketitle

\begingroup
\setlength{\abovedisplayskip}{1.5pt}
\setlength{\belowdisplayskip}{1.5pt}
\setlength{\abovedisplayshortskip}{1pt}
\setlength{\belowdisplayshortskip}{1pt}
\setlength{\jot}{0pt}

\begin{abstract}
 Vision-aided beam tracking reduces mmWave beam-training overhead by predicting beam indices from camera observations. Fixed top-$K$ selection, however, assigns the same candidate beam budget to all predictions, ignoring uncertainty variations across samples and horizons. This letter proposes confidence-guided adaptive candidate selection: predictor logits are temperature-calibrated, and the smallest top-ranked beam set whose cumulative probability reaches a target threshold is retained under a maximum budget. On DeepSense~6G Scenario~8, over five random seeds and at matched overhead, the policy improves candidate set accuracy by up to 3.80 percentage points, increases normalized received power by 0.32 percentage points, and reduces outage probability by 2.06 percentage points without retraining the predictor.
\end{abstract}


\section{Introduction}
\label{sec:introduction}

Reliable millimeter-wave (mmWave) vehicle-to-infrastructure communication requires the serving beam to remain aligned despite rapid changes in user position and channel geometry. Exhaustive search over a large beam codebook can maintain alignment but incurs substantial training overhead and latency. Vision-aided beam tracking reduces this cost by using camera observations at the base station (BS) to predict future beam indices \cite{jiang2022computer, alrabeiah2020millimeter}. To improve reliability, several highly ranked beams may be retained for subsequent refinement. However, assigning the same candidate count to every prediction ignores variations in uncertainty across samples and future horizons, potentially allocating too many beams to confident predictions and too few to uncertain ones.

Recent studies have improved vision-aided beam management through specialized visual representations, multi-camera sensing, multimodal fusion, robust temporal models, and proactive blockage handling \cite{charan2021vision}, while learning-based approaches have also reduced beam training overhead through adaptive selection and position-aided search \cite{ma2021deep,va2019online}. An encoder--decoder architecture using YOLOv4-based vehicle features and a GRU was developed to predict beam indices over multiple future horizons using the DeepSense~6G dataset \cite{jiang2022computer,alkhateeb2021deepsense6g}. Visual and positional information was combined with transmitter identification for beam prediction in real-world scenes containing multiple candidate vehicles \cite{charan2025camera}. Multiple camera views were exploited for proactive BS selection and efficient beam pair discovery \cite{lin2024multi}. Keypoint detection was used to extract the spatial distribution of propagation relevant scatterers, with the resulting environmental semantics employed for beam pair prediction \cite{wen2023vision}. Binary target mask sequences and a sequence-to-sequence model were used for occlusion-resilient multi-frame beam tracking \cite{park2025occlusion}. Decision level fusion of RGB images, GPS data, and received power measurements was applied together with beam misalignment detection and correction \cite{bian2024multimodal}.

These studies primarily improve sensing representations, target identification, multimodal robustness, BS selection, or the underlying beam predictor. In contrast, this letter studies a complementary resource-allocation problem at the prediction stage: how many candidate beams should be retained for each sample and prediction horizon. We adopt the predictor architecture of~\cite{jiang2022computer}, calibrate its output logits using temperature scaling~\cite{guo2017calibration}, and rank the resulting beam probabilities separately for each prediction. The proposed policy selects the smallest candidate beam set whose cumulative probability mass reaches a prescribed threshold. This rule is related to the cumulative-mass principle used in top-$p$, or nucleus, selection~\cite{holtzman2020curious}. Accordingly, we do not claim temperature scaling or cumulative-mass selection as new general machine-learning techniques. Rather, the contribution is their use for calibrated, sample- and horizon-dependent beam-budget allocation and systematic evaluation in vision-aided beam tracking.

The main contributions are: 1) confidence-calibrated adaptive candidate beam selection without modifying or retraining the underlying predictor; 2) cumulative-mass-based allocation subject to a deterministic maximum beam budget; and 3) a five-seed comparison with budget-matched fixed top-$K$ policies using candidate set accuracy, beam-training overhead, normalized received power, and outage probability.

Table~\ref{tab:related_work_comparison} highlights the key distinctions between the proposed adaptive candidate beam selection method and representative vision-aided beam management approaches.

\begin{table}[t]
\centering
\caption{Comparison with representative vision-aided beam management methods.}
\label{tab:related_work_comparison}

\begingroup
\scriptsize
\renewcommand{\arraystretch}{.6}
\setlength{\tabcolsep}{1.6pt}

\begin{tabularx}{\columnwidth}{
    |>{\raggedright\arraybackslash}X
    |>{\centering\arraybackslash}p{0.15\columnwidth}
    |>{\centering\arraybackslash}p{0.12\columnwidth}
    |>{\centering\arraybackslash}p{0.16\columnwidth}|
}
\hline
\textbf{Capability}
&
\textbf{This work}
&
\cite{jiang2022computer}
&
\cite{charan2025camera,lin2024multi,wen2023vision,
park2025occlusion,bian2024multimodal}
\\
\hline
Multi-horizon beam prediction
& $\checkmark$ & $\checkmark$ & $\triangle$ \\
\hline
Adaptive candidate count
& $\checkmark$ & -- & -- \\
\hline
Confidence calibration
& $\checkmark$ & -- & -- \\
\hline
Cumulative probability mass selection
& $\checkmark$ & -- & -- \\
\hline
Communication level  evaluation
& $\checkmark$ & $\checkmark$ & $\checkmark$ \\
\hline
Outage probability evaluation
& $\checkmark$ & -- & -- \\
\hline
\end{tabularx}

\vspace{0.4mm}
\parbox{\columnwidth}{%
\scriptsize
$\checkmark$: explicitly considered; $\triangle$: considered by a subset; --: not explicitly considered. Communication level evaluation includes received power, rate, throughput, or beam training overhead.%
}
\endgroup
\end{table}

\section{System Model and Problem Formulation}
\label{sec:system_model}

We consider a downlink millimeter-wave (mmWave) system in which a BS, equipped with an $N$-element antenna array and a co-located RGB camera, serves a high-mobility user equipment (UE), such as a vehicle. The BS employs a predefined beamforming codebook $\mathcal{F}=\{\mathbf{f}_1,\ldots,\mathbf{f}_C\}$, where $\mathbf{f}_c\in\mathbb{C}^{N\times 1}$ denotes the $c$-th beamforming vector and
$\mathcal{C}=\{1,\ldots,C\}$ is the corresponding beam index set.

\begin{figure}[H]
\centering

\resizebox{0.6\columnwidth}{!}{%
\begin{tikzpicture}[
    x=1cm,
    y=1cm,
    line cap=round,
    line join=round,
    >={Stealth[length=2.8mm,width=2.1mm]}
]


\definecolor{BSGreen}{RGB}{0,170,80}

\definecolor{CameraBlue}{RGB}{49,87,151}
\definecolor{CameraBlueDark}{RGB}{35,62,108}

\definecolor{BeamPink}{RGB}{224,72,119}
\definecolor{BeamRose}{RGB}{237,129,139}
\definecolor{BeamPeach}{RGB}{247,205,167}

\definecolor{CarBlue}{RGB}{25,121,218}
\definecolor{CarBlueDark}{RGB}{12,82,163}
\definecolor{CarBlueLight}{RGB}{74,166,235}

\definecolor{WindowCyan}{RGB}{164,229,245}
\definecolor{WindowCyanLight}{RGB}{218,247,252}

\definecolor{TyreDark}{RGB}{35,35,38}
\definecolor{WheelGray}{RGB}{202,205,208}
\definecolor{WheelDark}{RGB}{85,88,92}

\definecolor{LampYellow}{RGB}{255,210,70}
\definecolor{TailRed}{RGB}{224,36,55}

\definecolor{RoadGray}{RGB}{180,180,180}
\definecolor{TextBlack}{RGB}{0,0,0}


\draw[
    draw=BSGreen,
    line width=1.15pt
]
(1.15,3.15) rectangle (3.05,4.48);

\draw[
    draw=BSGreen,
    line width=0.82pt
]
(1.78,3.15) -- (1.78,4.48);

\draw[
    draw=BSGreen,
    line width=0.82pt
]
(2.42,3.15) -- (2.42,4.48);

\draw[
    draw=BSGreen,
    line width=0.82pt
]
(1.15,3.59) -- (3.05,3.59);

\draw[
    draw=BSGreen,
    line width=0.82pt
]
(1.15,4.03) -- (3.05,4.03);


\node[
    font=\bfseries\large,
    anchor=west,
    text=TextBlack
]
at (3.58,4.07)
{mmWave BS};


\fill[BSGreen]
(2.05,1.00) rectangle (2.17,3.15);


\fill[
    CameraBlue,
    rounded corners=1.6pt
]
(1.66,2.02) rectangle (2.56,2.66);

\fill[CameraBlue]
(1.86,2.66)
--
(1.99,2.86)
--
(2.25,2.86)
--
(2.38,2.66)
--
cycle;

\fill[white]
(2.11,2.34) circle (0.215);

\fill[CameraBlueDark]
(2.11,2.34) circle (0.145);

\fill[black]
(2.11,2.34) circle (0.078);

\fill[white]
(2.075,2.375) circle (0.030);

\fill[white]
(2.42,2.52) circle (0.030);


\node[
    font=\bfseries\large,
    anchor=east,
    text=TextBlack
]
at (1.43,2.34)
{RGB Camera};

%


\shade[
    left color=BeamPink,
    right color=BeamPeach,
    opacity=0.68,
    draw=none
]
(2.28,3.86)
.. controls (3.20,3.67) and (4.45,3.30) ..
(5.78,2.96)
.. controls (6.42,2.80) and (6.97,2.75) ..
(7.20,2.87)
.. controls (7.34,2.94) and (7.33,3.05) ..
(7.14,3.14)
.. controls (6.00,3.61) and (4.40,3.73) ..
(2.28,3.94)
-- cycle;


\shade[
    left color=BeamPink,
    right color=BeamRose!28!BeamPeach,
    opacity=0.34,
    draw=none
]
(2.34,3.885)
.. controls (3.25,3.72) and (4.38,3.42) ..
(5.60,3.11)
.. controls (6.13,2.98) and (6.56,2.96) ..
(6.81,3.03)
.. controls (6.93,3.07) and (6.92,3.13) ..
(6.76,3.20)
.. controls (5.82,3.53) and (4.34,3.67) ..
(2.34,3.925)
-- cycle;


\begin{scope}[shift={(6.08,0.88)}, scale=0.82]


\fill[
    RoadGray,
    opacity=0.30
]
(0.18,-0.05) ellipse (1.62 and 0.08);


\fill[
    CarBlue
]
(0.05,0.33)
--
(0.15,0.62)
--
(0.43,0.72)
--
(0.88,0.77)
--
(1.26,1.12)
--
(1.63,1.31)
--
(2.30,1.31)
--
(2.72,1.17)
--
(2.99,0.87)
--
(3.48,0.79)
--
(3.70,0.61)
--
(3.72,0.29)
--
(3.54,0.16)
--
(0.27,0.16)
--
cycle;

\fill[CarBlueDark]
(0.28,0.16)
rectangle
(3.52,0.27);


\fill[
    CarBlueLight
]
(0.92,0.78)
--
(1.30,1.12)
--
(1.64,1.26)
--
(2.28,1.26)
--
(2.67,1.13)
--
(2.92,0.82)
--
cycle;


\fill[
    WindowCyan
]
(2.14,1.19)
--
(2.61,1.08)
--
(2.82,0.83)
--
(2.13,0.83)
--
cycle;

\fill[
    WindowCyan
]
(1.63,1.19)
--
(2.04,1.19)
--
(2.04,0.83)
--
(1.25,0.83)
--
(1.47,1.09)
--
cycle;

\fill[
    WindowCyanLight,
    opacity=0.78
]
(1.52,1.14)
--
(1.66,1.17)
--
(1.46,0.85)
--
(1.32,0.85)
--
cycle;

\fill[
    WindowCyanLight,
    opacity=0.72
]
(2.31,1.14)
--
(2.45,1.10)
--
(2.28,0.85)
--
(2.16,0.85)
--
cycle;


\draw[
    draw=CarBlueDark,
    line width=0.75pt
]
(2.08,0.82) -- (2.08,1.22);

\draw[
    draw=CarBlueDark,
    line width=0.70pt
]
(1.21,0.81) -- (2.83,0.81);


\draw[
    draw=CarBlueDark,
    line width=0.65pt
]
(2.10,0.79) -- (2.10,0.27);

\draw[
    draw=CarBlueDark,
    line width=0.65pt
]
(2.94,0.78) -- (2.94,0.27);

\draw[
    draw=CarBlueDark,
    line width=0.75pt
]
(2.36,0.66) -- (2.55,0.66);

\draw[
    draw=CarBlueDark,
    line width=0.75pt
]
(3.08,0.64) -- (3.26,0.64);


\fill[
    LampYellow
]
(0.11,0.55)
--
(0.38,0.61)
--
(0.44,0.48)
--
(0.17,0.43)
--
cycle;


\fill[
    TailRed
]
(3.47,0.67)
--
(3.70,0.62)
--
(3.69,0.48)
--
(3.47,0.51)
--
cycle;


\fill[
    CarBlueDark
]
(0.04,0.30)
rectangle
(0.20,0.43);


\fill[TyreDark]
(0.77,0.15) circle (0.34);

\fill[WheelGray]
(0.77,0.15) circle (0.235);

\fill[WheelDark]
(0.77,0.15) circle (0.085);

\foreach \a in {0,60,...,300}{
    \draw[
        WheelDark,
        line width=0.55pt
    ]
    (0.77,0.15)
    --
    ++(\a:0.20);
}

\fill[TyreDark]
(3.05,0.15) circle (0.34);

\fill[WheelGray]
(3.05,0.15) circle (0.235);

\fill[WheelDark]
(3.05,0.15) circle (0.085);

\foreach \a in {0,60,...,300}{
    \draw[
        WheelDark,
        line width=0.55pt
    ]
    (3.05,0.15)
    --
    ++(\a:0.20);
}


\draw[
    draw=CarBlueDark,
    line width=0.85pt
]
(0.05,0.33)
--
(0.15,0.62)
--
(0.43,0.72)
--
(0.88,0.77)
--
(1.26,1.12)
--
(1.63,1.31)
--
(2.30,1.31)
--
(2.72,1.17)
--
(2.99,0.87)
--
(3.48,0.79)
--
(3.70,0.61)
--
(3.72,0.29);

\end{scope}


\draw[
    -{Stealth[length=3.2mm,width=2.4mm]},
    line width=0.85pt,
    draw=black
]
(6.22,1.60) -- (4.55,1.60);


\node[
    font=\bfseries\large,
    text=TextBlack
]
at (7.38,0.45)
{Mobile UE};

\end{tikzpicture}%
}
\caption{Illustration of an RGB camera-aided mmWave beam tracking setup.}
\label{fig:mmwave_camera_setup}
\end{figure}
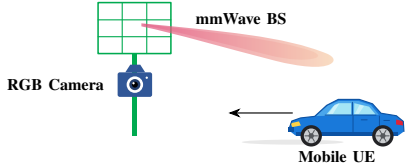

At time step $t$, the signal received by the UE is
$y[t]
=
\mathbf{h}^{H}[t]\mathbf{f}[t]s[t]+n[t]$,
where $\mathbf{h}[t]\in\mathbb{C}^{N\times1}$ denotes the downlink
channel, $\mathbf{f}[t]\in\mathcal{F}$ is the selected beamforming
vector, $s[t]$ is the transmitted symbol, and $n[t]$ denotes additive
noise. The optimal beam index is defined as
$c^{\star}[t]
=
\underset{c\in\mathcal{C}}{\arg\max}
\left|
\mathbf{h}^{H}[t]\mathbf{f}_{c}
\right|^{2}$.
Determining $c^{\star}[t]$ through exhaustive beam sweeping requires
testing all $C$ codebook beams. Repeating this procedure frequently in a
high-mobility environment can incur considerable beam training
overhead. The co-located camera provides visual observations of the UE
position and motion that can be used to predict future beam indices, as
illustrated in Fig.~\ref{fig:mmwave_camera_setup}.

Let
$\mathbf{X}[t]\in
\mathbb{R}^{H_{\mathrm{img}}\times W_{\mathrm{img}}\times3}$
denote the RGB image acquired at time step $t$. Each input sample
contains $H_{\mathrm{obs}}$ consecutive RGB frames and is used to predict
the optimal beam indices over the following $H_{\mathrm{pred}}$ time
steps.
For sample $i$, the observation sequence is
$\mathcal{X}_{i}
=
\{\mathbf{X}_{i,1},\ldots,\mathbf{X}_{i,H_{\mathrm{obs}}}\}$.
The corresponding ground truth optimal beam indices are
$\{c_{i,1}^{\star},\ldots,c_{i,H_{\mathrm{pred}}}^{\star}\}$, where
$c_{i,r}^{\star}\in\mathcal{C}$.
For each sample $i$ and prediction step
$r\in\{1,\ldots,H_{\mathrm{pred}}\}$, the beam tracking method constructs
a candidate set $\mathcal{S}_{i,r}\subseteq\mathcal{C}$. The objective is
to include the optimal beam $c_{i,r}^{\star}$ while keeping the candidate
count $k_{i,r}=|\mathcal{S}_{i,r}|$ small. This balances beam tracking
reliability and beam training overhead. The adopted predictor and the
proposed candidate selection method are described in
Section~\ref{sec:proposed_method}.

\section{Proposed Adaptive Candidate Beam Selection}
\label{sec:proposed_method}

The beam tracking task is formulated as a multiclass classification
problem over the beam index set
$\mathcal{C}=\{1,\ldots,C\}$, with $C=64$ for DeepSense~6G
Scenario~8 \cite{alkhateeb2021deepsense6g}. We adopt the vision-aided beam tracking architecture in \cite{jiang2022computer}, shown in Fig.~\ref{fig:encoder_decoder_beam_prediction}, as the underlying beam predictor. YOLOv4 extracts vehicle features from the RGB image sequence, which are processed by the encoder--decoder GRU architecture to generate beam logits for multiple future prediction horizons. The predictor architecture and its training procedure remain unchanged. The contribution of this letter is introduced after prediction, as illustrated in Fig.~\ref{fig:proposed_pipeline}. The resulting beam logits are calibrated using temperature scaling, and the calibrated probabilities are subsequently used for adaptive candidate beam selection.

\begin{figure}[t]
\centering
\resizebox{0.72\columnwidth}{!}{%
\begin{tikzpicture}[
    x=0.94cm,
    y=0.94cm,
    line cap=round,
    line join=round,
    >={Stealth[length=2.5mm,width=1.8mm]}
]


\definecolor{EncMain}{RGB}{49,87,151}
\definecolor{EncFill}{RGB}{240,245,251}
\definecolor{EncFillAlt}{RGB}{246,248,252}

\definecolor{DecMain}{RGB}{170,72,95}
\definecolor{DecFill}{RGB}{252,242,245}
\definecolor{DecFillAlt}{RGB}{253,247,249}

\definecolor{FeatureMain}{RGB}{177,143,45}
\definecolor{FeatureFill}{RGB}{253,249,235}

\definecolor{TextDark}{RGB}{0,0,0}

\definecolor{BorderGray}{RGB}{105,110,117}
\definecolor{OuterGray}{RGB}{125,130,137}
\definecolor{ArrowGray}{RGB}{35,38,42}
\definecolor{SoftGray}{RGB}{249,249,249}


\def\encL{0.10}
\def\encR{3.30}

\def\decL{4.90}
\def\decR{8.10}

\def\outerB{0.88}
\def\outerT{6.05}


\fill[SoftGray]
    (\encL,\outerB) rectangle (\encR,\outerT);

\fill[SoftGray]
    (\decL,\outerB) rectangle (\decR,\outerT);

\draw[
    dashed,
    dash pattern=on 4pt off 2.7pt,
    line width=0.55pt,
    draw=OuterGray
]
    (\encL,\outerB) rectangle (\encR,\outerT);

\draw[
    dashed,
    dash pattern=on 4pt off 2.7pt,
    line width=0.55pt,
    draw=OuterGray
]
    (\decL,\outerB) rectangle (\decR,\outerT);


\node[
    font=\bfseries\large,
    text=TextDark
] at (1.70,5.72)
{Encoder};

\node[
    font=\bfseries\large,
    text=TextDark
] at (6.50,5.72)
{Decoder};

\fill[EncMain]
    (\encL,5.38) rectangle (\encR,5.44);

\fill[DecMain]
    (\decL,5.38) rectangle (\decR,5.44);



\fill[EncFill]
    (0.28,4.60) rectangle (3.12,5.13);

\draw[
    line width=0.48pt,
    draw=EncMain
]
    (0.28,4.60) rectangle (3.12,5.13);

\fill[EncMain]
    (0.28,4.60) rectangle (0.34,5.13);

\node[
    font=\bfseries\fontsize{6.7}{7.2}\selectfont,
    text=TextDark
] at (1.70,4.865)
{GRU encoder (64)};


\fill[EncFillAlt]
    (0.28,3.47) rectangle (3.12,4.19);

\draw[
    line width=0.48pt,
    draw=EncMain
]
    (0.28,3.47) rectangle (3.12,4.19);

\fill[EncMain]
    (0.28,3.47) rectangle (0.34,4.19);

\node[
    font=\bfseries\fontsize{6.45}{7.0}\selectfont,
    align=center,
    text=TextDark
] at (1.70,3.83)
{Fully connected\\embedding};


\fill[FeatureFill]
    (0.28,2.43) rectangle (3.12,3.06);

\draw[
    line width=0.48pt,
    draw=FeatureMain
]
    (0.28,2.43) rectangle (3.12,3.06);

\fill[FeatureMain]
    (0.28,2.43) rectangle (0.34,3.06);

\node[
    font=\bfseries\fontsize{6.35}{6.9}\selectfont,
    text=TextDark
] at (1.70,2.745)
{YOLOv4 feature extraction};


\fill[white]
    (0.28,1.15) rectangle (3.12,2.08);

\draw[
    line width=0.48pt,
    draw=BorderGray
]
    (0.28,1.15) rectangle (3.12,2.08);

\fill[EncMain]
    (0.28,1.15) rectangle (0.34,2.08);


\begin{scope}[shift={(0.55,1.37)}]

    \fill[
        EncMain,
        rounded corners=1.1pt
    ]
        (0,0) rectangle (0.82,0.50);

    \fill[EncMain]
        (0.18,0.50)
        -- (0.31,0.66)
        -- (0.55,0.66)
        -- (0.67,0.50)
        -- cycle;

    \fill[white]
        (0.41,0.25) circle (0.165);

    \fill[EncMain!72!black]
        (0.41,0.25) circle (0.105);

    \fill[white]
        (0.41,0.25) circle (0.046);

\end{scope}

\node[
    font=\bfseries\fontsize{6.35}{6.90}\selectfont,
    align=center,
    text width=1.34cm,
    text=TextDark
] at (2.23,1.615)
{Past RGB\\image\\sequence};


\draw[
    ->,
    draw=ArrowGray,
    line width=0.76pt
]
    (1.70,2.11) -- (1.70,2.38);

\draw[
    ->,
    draw=ArrowGray,
    line width=0.76pt
]
    (1.70,3.09) -- (1.70,3.42);

\draw[
    ->,
    draw=ArrowGray,
    line width=0.76pt
]
    (1.70,4.22) -- (1.70,4.55);



\fill[DecFill]
    (5.08,4.55) rectangle (7.92,5.17);

\draw[
    line width=0.48pt,
    draw=DecMain
]
    (5.08,4.55) rectangle (7.92,5.17);

\fill[DecMain]
    (5.08,4.55) rectangle (5.14,5.17);

\node[
    font=\fontsize{6.35}{6.90}\selectfont,
    align=center,
    text width=2.30cm,
    text=TextDark
] at (6.50,4.86)
{Beam logits over the\\codebook};


\fill[DecFillAlt]
    (5.08,3.36) rectangle (7.92,4.14);

\draw[
    line width=0.48pt,
    draw=DecMain
]
    (5.08,3.36) rectangle (7.92,4.14);

\fill[DecMain]
    (5.08,3.36) rectangle (5.14,4.14);

\node[
    font=\bfseries\fontsize{6.35}{6.90}\selectfont,
    align=center,
    text width=2.45cm,
    text=TextDark
] at (6.50,3.75)
{Fully connected classifier};


\fill[FeatureFill]
    (5.08,2.33) rectangle (7.92,3.02);

\draw[
    line width=0.48pt,
    draw=FeatureMain
]
    (5.08,2.33) rectangle (7.92,3.02);

\fill[FeatureMain]
    (5.08,2.33) rectangle (5.14,3.02);

\node[
    font=\bfseries\fontsize{6.55}{7.05}\selectfont,
    text=TextDark
] at (6.50,2.675)
{GRU decoder (64)};


\fill[DecFill]
    (5.08,1.40) rectangle (7.92,2.12);

\draw[
    line width=0.48pt,
    draw=DecMain
]
    (5.08,1.40) rectangle (7.92,2.12);

\fill[DecMain]
    (5.08,1.40) rectangle (5.14,2.12);

\node[
    font=\fontsize{6.20}{6.75}\selectfont,
    align=center,
    text width=2.38cm,
    text=TextDark
] at (6.50,1.76)
{Embedding of zero-vector\\padding};


\fill[white]
    (5.08,0.94) rectangle (7.92,1.27);

\draw[
    line width=0.48pt,
    draw=BorderGray
]
    (5.08,0.94) rectangle (7.92,1.27);

\node[
    font=\fontsize{6.05}{6.55}\selectfont,
    text=TextDark
] at (6.50,1.105)
{Zero-vector decoder input};


\draw[
    ->,
    draw=ArrowGray,
    line width=0.76pt
]
    (6.50,1.38) -- (6.50,1.44);

\draw[
    ->,
    draw=ArrowGray,
    line width=0.76pt
]
    (6.50,2.20) -- (6.50,2.35);

\draw[
    ->,
    draw=ArrowGray,
    line width=0.76pt
]
    (6.50,3.06) -- (6.50,3.41);

\draw[
    ->,
    draw=ArrowGray,
    line width=0.76pt
]
    (6.50,4.18) -- (6.50,4.50);


\draw[
    ->,
    draw=ArrowGray,
    line width=0.84pt
]
    (3.30,3.30) -- (4.82,3.30);

\node[
    font=\fontsize{5.75}{6.20}\selectfont,
    align=center,
    text width=1.12cm,
    text=TextDark,
    fill=white,
    inner sep=0.8pt
]
at (4.06,3.77)
{Encoded hidden\\
state from past\\
observations};


\draw[
    ->,
    draw=TextDark,
    line width=0.82pt
]
    (0.12,0.60) -- (7.18,0.60);

\node[
    font=\bfseries\fontsize{6.45}{6.95}\selectfont,
    anchor=west,
    text=TextDark
]
at (7.23,0.60)
{Time};


\node[
    font=\bfseries\fontsize{6.35}{6.90}\selectfont,
    text=EncMain
]
at (1.40,0.12)
{Past observations};

\node[
    font=\bfseries\fontsize{6.35}{6.90}\selectfont,
    text=DecMain
]
at (6.14,0.12)
{Future prediction steps};

\end{tikzpicture}
}
\caption{Vision-aided beam tracking architecture adopted from \cite{jiang2022computer} for multi-horizon beam prediction.}
\label{fig:encoder_decoder_beam_prediction}
\end{figure}
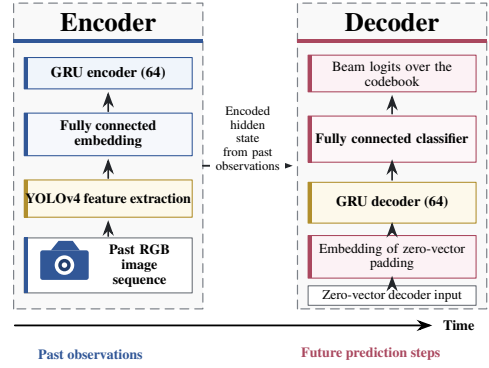

\subsection{Confidence Calibration}
\label{subsec:confidence_calibration}

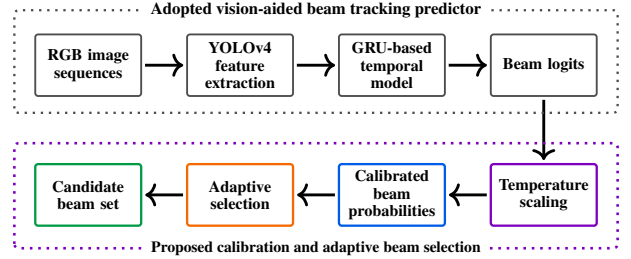
\begin{figure}[t]
\centering

\definecolor{VioletLine}{RGB}{126,0,190}
\definecolor{BlueLine}{RGB}{0,92,230}
\definecolor{OrangeLine}{RGB}{245,105,0}
\definecolor{GreenLine}{RGB}{0,155,70}
\resizebox{0.92\columnwidth}{!}{%
\begin{tikzpicture}[
    every node/.style={
        font=\fontsize{5.6}{6.1}\selectfont,
        align=center,
        text=black
    },
    topbox/.style={
        draw=black!70,
        fill=white,
        rounded corners=1pt,
        line width=0.65pt,
        minimum height=7.5mm,
        text width=1.15cm,
        inner xsep=2pt,
        inner ysep=2pt
    },
    bottombox/.style={
        fill=white,
        rounded corners=1pt,
        line width=0.8pt,
        minimum height=7.5mm,
        text width=1.15cm,
        inner xsep=2pt,
        inner ysep=2pt
    },
    flow/.style={
        ->,
        draw=black,
        line width=1.0pt,
        shorten <=1pt,
        shorten >=1pt
    }
]

\node[topbox] (rgb) at (0.80,0.80)
{\textbf{RGB image}\\\textbf{sequences}};

\node[topbox] (yolo) at (2.65,0.80)
{\textbf{YOLOv4 feature extraction}};

\node[topbox] (gru) at (4.50,0.80)
{\textbf{GRU-based temporal model}};

\node[topbox] (logits) at (6.35,0.80)
{\textbf{Beam logits}};

\draw[flow] (rgb.east) -- (yolo.west);
\draw[flow] (yolo.east) -- (gru.west);
\draw[flow] (gru.east) -- (logits.west);

\node[
    bottombox,
    draw=GreenLine
] (candidate) at (0.80,-0.78)
{\textbf{Candidate}\\\textbf{beam set}};

\node[
    bottombox,
    draw=OrangeLine
] (adaptive) at (2.65,-0.78)
{\textbf{Adaptive}\\\textbf{selection}};

\node[
    bottombox,
    draw=BlueLine
] (probabilities) at (4.50,-0.78)
{\textbf{Calibrated beam}\\\textbf{probabilities}};

\node[
    bottombox,
    draw=VioletLine
] (temperature) at (6.35,-0.78)
{\textbf{Temperature}\\\textbf{scaling}};

\draw[flow] (temperature.west) -- (probabilities.east);
\draw[flow] (probabilities.west) -- (adaptive.east);
\draw[flow] (adaptive.west) -- (candidate.east);
\draw[flow] (logits.south) -- (temperature.north);

\draw[
    dotted,
    draw=black!70,
    line width=0.8pt,
    rounded corners=1.5pt
]
(-0.10,0.22) rectangle (7.25,1.47);

\node[
    fill=white,
    inner xsep=3pt,
    font=\fontsize{5.8}{6.2}\selectfont\bfseries
] at (3.575,1.47)
{Adopted vision-aided beam tracking predictor};

\draw[
    dotted,
    draw=VioletLine,
    line width=0.9pt,
    rounded corners=1.5pt
]
(-0.10,-1.43) rectangle (7.25,-0.15);

\node[
    fill=white,
    inner xsep=3pt,
    inner ysep=0.5pt,
    font=\fontsize{5.4}{5.9}\selectfont\bfseries
] at (3.575,-1.43)
{Proposed calibration and adaptive beam selection};

\end{tikzpicture}
}
\caption{Proposed confidence-guided adaptive candidate beam selection framework applied to the output of the adopted beam predictor.}
\label{fig:proposed_pipeline}
\end{figure}

For sample $i$, prediction step
$r\in\{1,\ldots,H_{\mathrm{pred}}\}$, and random seed
$s\in\{1,\ldots,S\}$, the adopted network of
\cite{jiang2022computer} produces the logit vector
$\mathbf{z}_{i,r}^{(s)}
=[z_{i,r,1}^{(s)},\ldots,z_{i,r,C}^{(s)}]^{\top}
\in\mathbb{R}^{C}$. Temperature scaling
\cite{guo2017calibration} converts these logits into
\begin{equation}
p_{i,r,c}^{(s)}(T)
=
\frac{\exp\!\left(z_{i,r,c}^{(s)}/T\right)}
{\sum_{\ell=1}^{C}
\exp\!\left(z_{i,r,\ell}^{(s)}/T\right)},
\quad
T>0,\quad c\in\mathcal{C}.
\label{eq:temperature_scaling}
\end{equation}
A temperature below one sharpens the probability distribution, whereas
a temperature above one softens it, without changing the beam ordering.

The sequence samples are divided into training, calibration, and test
sets containing approximately $72\%$, $8\%$, and $20\%$ of the data,
respectively. Let $\mathcal{I}_{\mathrm{cal}}$ denote the set of
calibration samples and define
$\mathcal{D}_{\mathrm{cal}}
=
\{(i,r):i\in\mathcal{I}_{\mathrm{cal}},
r\in\{1,\ldots,H_{\mathrm{pred}}\}\}$.
Thus,
$N_{\mathrm{cal}}
=
|\mathcal{D}_{\mathrm{cal}}|
=
|\mathcal{I}_{\mathrm{cal}}|H_{\mathrm{pred}}$.

For each seed, a single global temperature is fitted jointly over all
calibration samples and prediction steps by minimizing the negative log
likelihood:
\begin{equation}
T_s^{\star}
=
\underset{T>0}{\arg\min}
\left[
-\frac{1}{N_{\mathrm{cal}}}
\sum_{(i,r)\in\mathcal{D}_{\mathrm{cal}}}
\log
p_{i,r,c_{i,r}^{\star}}^{(s)}(T)
\right],
\label{eq:temperature_optimization}
\end{equation}
where $c_{i,r}^{\star}\in\mathcal{C}$ denotes the ground truth optimal
beam defined in Section~\ref{sec:system_model}. The uncalibrated and
calibrated probabilities are given by
$p_{i,r,c}^{(s)}(1)$ and
$p_{i,r,c}^{(s)}(T_s^{\star})$, respectively.

The calibration and evaluation procedures are performed independently
for $S=5$ random seeds, after which the resulting performance metrics
are aggregated.  {Since scaling the logits by any $T>0$ preserves their
ordering, temperature scaling does not alter fixed top- $K$ candidate
sets. It does, however, change the probability concentration and can
therefore affect the number of beams required to reach a prescribed
cumulative probability mass.}

For compactness, the seed index is omitted in the following subsection.
Unless otherwise stated, $p_{i,r,c}$ denotes the calibrated probability
$p_{i,r,c}^{(s)}(T_s^{\star})$ for a fixed seed $s$.
\subsection{Adaptive Candidate Beam Selection}
\label{subsec:adaptive_selection}

Let $\pi_{i,r}$ be a permutation of $\mathcal{C}$ that orders the
calibrated probabilities as
$p_{i,r,\pi_{i,r}(1)}
\geq \cdots \geq
p_{i,r,\pi_{i,r}(C)}$.
For notational convenience, let
$p_{i,r}^{(m)}=p_{i,r,\pi_{i,r}(m)}$
and
$F_{i,r}(k)=\sum_{m=1}^{k}p_{i,r}^{(m)}$.
 {For a cumulative mass threshold $\tau\in(0,1)$, the adaptive candidate
count is
\begin{equation*}
k_{i,r}^{\mathrm{adp}}(\tau)
=
\min\!\left\{
K_{\max},
\min\!\left\{
k\in\{1,\ldots,C\}:
F_{i,r}(k)\geq\tau
\right\}
\right\}.
\end{equation*}}
Moreover, the corresponding candidate set is
$\mathcal{S}_{i,r}^{\mathrm{adp}}(\tau)
=
\left\{
\pi_{i,r}(1),\ldots,
\pi_{i,r}\!\left(k_{i,r}^{\mathrm{adp}}(\tau)\right)
\right\}$.
For comparison, fixed top-$K$ selection uses
$\mathcal{S}_{i,r}^{\mathrm{fix}}(K)
=
\{\pi_{i,r}(1),\ldots,\pi_{i,r}(K)\}$ and therefore assigns the same candidate count to every prediction.

Prior work on DeepSense~6G Scenario~8 \cite{alkhateeb2021deepsense6g} with a $64$ beam codebook reported a top-$5$ accuracy of $98.95\%$, indicating that the optimal beam is typically contained within a small set of highly ranked candidate beams \cite{jiang2022computer}.  {We set $K_{\max}=8$ to provide additional flexibility for uncertain predictions while limiting the maximum beam-training overhead to $100K_{\max}/C=12.5\%$.}

 {
The threshold $\tau$ determines the operating point between candidate
count and reliability. In this work, we evaluate
$\tau\in
\{0.50,0.60,0.70,0.80,0.90,0.93,0.95,0.97,0.98,0.99\}$.}

\subsection{Beam Tracking Performance Metrics}
\label{subsec:performance_metrics}

For a fixed seed, let
$\mathcal{S}_{i,r}\subseteq\mathcal{C}$ denote the candidate set for test
sample $i$ and prediction step $r$, with
$k_{i,r}=|\mathcal{S}_{i,r}|$. The seed index is omitted, and
$N_{\mathrm{eval}}=N_{\mathrm{test}}H_{\mathrm{pred}}$.
 {The average candidate count and beam training overhead are
\begin{equation}
\bar{k}
=
\frac{1}{N_{\mathrm{eval}}}
\sum_{i=1}^{N_{\mathrm{test}}}
\sum_{r=1}^{H_{\mathrm{pred}}}
k_{i,r},
\qquad
\mathrm{OH}
=
100\frac{\bar{k}}{C}.
\label{eq:beam_overhead}
\end{equation}
where $k_{i,r}=K$ for fixed top-$K$ selection and
$k_{i,r}=k_{i,r}^{\mathrm{adp}}(\tau)$ for the adaptive policy.}
Moreover, the candidate set accuracy is
$\mathrm{ACC}
=
\frac{1}{N_{\mathrm{eval}}}
\sum_{i=1}^{N_{\mathrm{test}}}
\sum_{r=1}^{H_{\mathrm{pred}}}
\mathbf{1}
\left\{
c_{i,r}^{\star}\in\mathcal{S}_{i,r}
\right\}$.

Let $P_{i,r}(c)$ denote the measured received power in decibels and
$G_{i,r}(c)=10^{P_{i,r}(c)/10}$ its linear value. The normalized
received power and its average are
$R_{i,r}
=
\frac{
\max_{c\in\mathcal{S}_{i,r}}G_{i,r}(c)
}{
G_{i,r}(c_{i,r}^{\star})
},$ and
$\bar{R}
=
\frac{1}{N_{\mathrm{eval}}}
\sum_{i=1}^{N_{\mathrm{test}}}
\sum_{r=1}^{H_{\mathrm{pred}}}
R_{i,r}$,
respectively. 
Since $c_{i,r}^{\star}$ is the optimal beam,
$0\leq R_{i,r}\leq1$.

 {An outage occurs when $R_{i,r}<\rho_{\min}$. We set $\rho_{\min}=0.95$, ensuring at least $95\%$ of the optimal linear received power with a maximum loss of approximately $0.22$~dB, consistent with prior mmWave beamforming analysis using a $95\%$ maximum power criterion.} The outage probability is
\begin{equation}
\mathrm{OP}
=
\frac{1}{N_{\mathrm{eval}}}
\sum_{i=1}^{N_{\mathrm{test}}}
\sum_{r=1}^{H_{\mathrm{pred}}}
\mathbf{1}
\left\{
R_{i,r}<\rho_{\min}
\right\}.
\label{eq:outage_probability}
\end{equation}
All metrics are computed for each seed and then
aggregated. The quantities $\mathrm{ACC}$, $\bar{R}$, and
$\mathrm{OP}$ are reported as percentages.

\subsection{Budget Matched Fixed Top-\texorpdfstring{$K$}{K} Comparison}
\label{subsec:budget_matching}

 {The adaptive policy generally yields a noninteger average candidate
count. To compare the adaptive and fixed policies at the same average
overhead, the fixed top-$K$ results are linearly interpolated between the
evaluated budgets
$\mathcal{K}_{\mathrm{fix}}=\{1,2,3,5,8\}$.}

For seed $s$ and threshold $\tau$, let
$\bar{k}_{s}^{\mathrm{adp}}(\tau)$ denote the average adaptive candidate
count over all test prediction instances. When this value lies strictly
between adjacent budgets
$K_a,K_b\in\mathcal{K}_{\mathrm{fix}}$, with
$K_a<\bar{k}_{s}^{\mathrm{adp}}(\tau)<K_b$, define
$\lambda_s(\tau)
=
\frac{\bar{k}_{s}^{\mathrm{adp}}(\tau)-K_a}
{K_b-K_a}$.
For
$\mathcal{M}\in\{\mathrm{ACC},\bar{R},\mathrm{OP}\}$, the matched fixed
value is
\begin{equation}
\mathcal{M}_{s}^{\mathrm{match}}(\tau)
=
\left[1-\lambda_s(\tau)\right]
\mathcal{M}_{s}^{\mathrm{fix}}(K_a)
+
\lambda_s(\tau)
\mathcal{M}_{s}^{\mathrm{fix}}(K_b).
\label{eq:budget_matching}
\end{equation}

If $\bar{k}_{s}^{\mathrm{adp}}(\tau)$ equals an evaluated fixed budget,
the corresponding fixed result is used directly.  {Matching is performed
independently for each seed before aggregation. Since
$\mathrm{OH}=100\bar{k}/C$, equal average candidate counts also imply
equal average beam training overhead.}

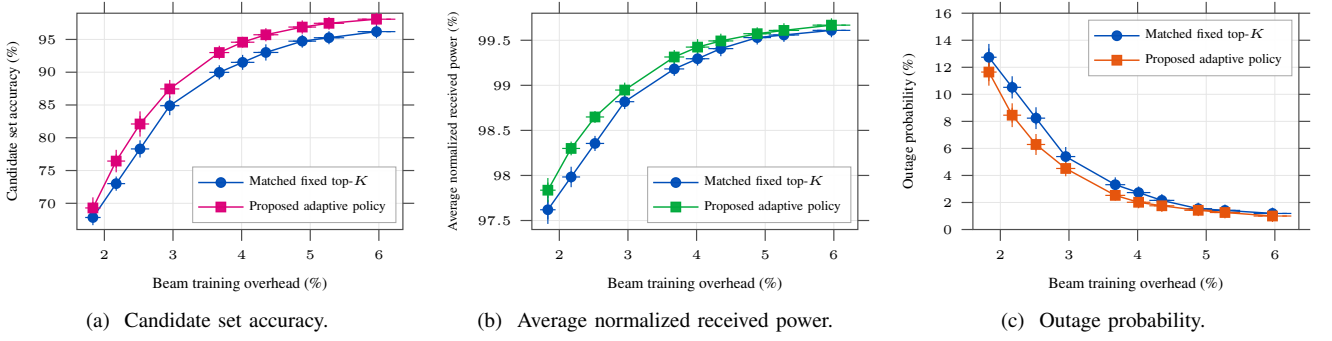
\begin{figure*}[t]
\centering

\makebox[\textwidth][c]{%
\subfloat[
Candidate set accuracy.
\label{fig:overhead_topk_accuracy_tradeoff}]
{%
\begin{tikzpicture}
\begin{axis}[
    resultaxis,
    width=0.325\textwidth,
    height=0.245\textwidth,
    xlabel={Beam training overhead (\%)},
    ylabel={Candidate set accuracy (\%)},
    xlabel style={
        font=\fontsize{5.8}{6.2}\selectfont
    },
    ylabel style={
        font=\fontsize{5.5}{5.9}\selectfont
    },
    tick label style={
        font=\fontsize{5.3}{5.7}\selectfont
    },
    ymin=66,
    ymax=99,
    ytick={70,75,80,85,90,95},
    legend style={
        font=\fontsize{5.0}{5.4}\selectfont,
        at={(0.98,0.03)},
        anchor=south east,
        inner xsep=1.8pt,
        inner ysep=1.4pt,
        row sep=0.3pt
    }
]

\addplot+[
    fixedcurve,
    /pgfplots/.cd,
    solid,
    line width=0.55pt,
    draw=FixedBlue,
    color=FixedBlue,
    mark options={
        draw=FixedBlue,
        fill=FixedBlue
    }
]
table[
    row sep=\\,
    x=overhead,
    x error=xerr,
    y=accuracy,
    y error=yerr
]{
overhead xerr     accuracy  yerr\\
1.832993 0.046676 67.834890 0.482392\\
2.172206 0.064673 73.001775 0.398005\\
2.519678 0.067474 78.297614 0.624058\\
2.952590 0.073672 84.892491 0.759014\\
3.674653 0.079191 89.982007 0.388698\\
4.014682 0.105772 91.503804 0.451167\\
4.353385 0.116184 93.014696 0.552272\\
4.883462 0.136272 94.736991 0.302622\\
5.272940 0.155646 95.260620 0.305813\\
5.965334 0.209306 96.190677 0.323606\\
};
\addlegendentry{Matched fixed top-$K$}

\addplot+[
    adaptivecurve,
    /pgfplots/.cd,
    solid,
    line width=0.55pt,
    draw=AdaptiveMagenta,
    color=AdaptiveMagenta,
    mark options={
        draw=AdaptiveMagenta,
        fill=AdaptiveMagenta
    }
]
table[
    row sep=\\,
    x=overhead,
    x error=xerr,
    y=accuracy,
    y error=yerr
]{
overhead xerr     accuracy  yerr\\
1.832993 0.046676 69.292007 0.926096\\
2.172206 0.064673 76.443719 1.024750\\
2.519678 0.067474 82.101142 1.238645\\
2.952590 0.073672 87.451876 0.680713\\
3.674653 0.079191 92.998369 0.351851\\
4.014682 0.105772 94.577488 0.135311\\
4.353385 0.116184 95.712887 0.311749\\
4.883462 0.136272 96.887439 0.326753\\
5.272940 0.155646 97.461664 0.274334\\
5.965334 0.209306 98.088091 0.263649\\
};
\addlegendentry{Proposed adaptive policy}

\end{axis}
\end{tikzpicture}
}
\hspace{0.004\textwidth}
\subfloat[
Average normalized received power.
\label{fig:budget_matched_normalized_power}]
{%
\begin{tikzpicture}
\begin{axis}[
    resultaxis,
    width=0.325\textwidth,
    height=0.245\textwidth,
    xlabel={Beam training overhead (\%)},
    ylabel={Average normalized received power (\%)},
    xlabel style={
        font=\fontsize{5.8}{6.2}\selectfont
    },
    ylabel style={
        font=\fontsize{5.0}{5.4}\selectfont
    },
    tick label style={
        font=\fontsize{5.3}{5.7}\selectfont
    },
    ymin=97.4,
    ymax=99.8,
    ytick={97.5,98,98.5,99,99.5},
    legend style={
        font=\fontsize{5.0}{5.4}\selectfont,
        at={(0.98,0.03)},
        anchor=south east,
        inner xsep=1.8pt,
        inner ysep=1.4pt,
        row sep=0.3pt
    }
]

\addplot+[
    fixedcurve,
    /pgfplots/.cd,
    solid,
    line width=0.55pt,
    draw=FixedBlue,
    color=FixedBlue,
    mark options={
        draw=FixedBlue,
        fill=FixedBlue
    }
]
table[
    row sep=\\,
    x=overhead,
    x error=xerr,
    y=power,
    y error=yerr
]{
overhead xerr     power      yerr\\
1.832993 0.046676 97.619458  0.107239\\
2.172206 0.064673 97.983139  0.063467\\
2.519678 0.067474 98.355107  0.033654\\
2.952590 0.073672 98.818389  0.031636\\
3.674653 0.079191 99.181889  0.027065\\
4.014682 0.105772 99.294999  0.025647\\
4.353385 0.116184 99.407318  0.034022\\
4.883462 0.136272 99.529717  0.026554\\
5.272940 0.155646 99.558595  0.026217\\
5.965334 0.209306 99.609900  0.026573\\
};
\addlegendentry{Matched fixed top-$K$}

\addplot+[
    adaptivecurve,
    /pgfplots/.cd,
    solid,
    line width=0.55pt,
    draw=AdaptiveGreen,
    color=AdaptiveGreen,
    mark options={
        draw=AdaptiveGreen,
        fill=AdaptiveGreen
    }
]
table[
    row sep=\\,
    x=overhead,
    x error=xerr,
    y=power,
    y error=yerr
]{
overhead xerr     power      yerr\\
1.832993 0.046676 97.836399  0.084865\\
2.172206 0.064673 98.299605  0.026049\\
2.519678 0.067474 98.648965  0.015739\\
2.952590 0.073672 98.947603  0.032374\\
3.674653 0.079191 99.314398  0.026956\\
4.014682 0.105772 99.424219  0.037841\\
4.353385 0.116184 99.492562  0.030250\\
4.883462 0.136272 99.572818  0.020509\\
5.272940 0.155646 99.609246  0.029324\\
5.965334 0.209306 99.667835  0.025493\\
};
\addlegendentry{Proposed adaptive policy}

\end{axis}
\end{tikzpicture}
}
\hspace{0.004\textwidth}
\subfloat[
Outage probability.
\label{fig:overhead_outage_tradeoff}]
{%
\begin{tikzpicture}
\begin{axis}[
    resultaxis,
    width=0.325\textwidth,
    height=0.245\textwidth,
    xlabel={Beam training overhead (\%)},
    ylabel={Outage probability (\%)},
    xlabel style={
        font=\fontsize{5.8}{6.2}\selectfont
    },
    ylabel style={
        font=\fontsize{5.5}{5.9}\selectfont
    },
    tick label style={
        font=\fontsize{5.3}{5.7}\selectfont
    },
    ymin=0,
    ymax=16,
    ytick={0,2,4,6,8,10,12,14,16},
    legend style={
        font=\fontsize{5.0}{5.4}\selectfont,
        at={(0.98,0.97)},
        anchor=north east,
        inner xsep=1.8pt,
        inner ysep=1.4pt,
        row sep=0.3pt
    }
]

\addplot+[
    fixedcurve,
    /pgfplots/.cd,
    solid,
    line width=0.55pt,
    draw=FixedBlue,
    color=FixedBlue,
    mark options={
        draw=FixedBlue,
        fill=FixedBlue
    }
]
table[
    row sep=\\,
    x=overhead,
    x error=xerr,
    y=outage,
    y error=yerr
]{
overhead xerr     outage    yerr\\
1.832993 0.046676 12.743440 0.651323\\
2.172206 0.064673 10.516821 0.486878\\
2.519678 0.067474  8.237993 0.472418\\
2.952590 0.073672  5.397003 0.372243\\
3.674653 0.079191  3.315971 0.208295\\
4.014682 0.105772  2.732685 0.114307\\
4.353385 0.116184  2.155690 0.129393\\
4.883462 0.136272  1.542592 0.104885\\
5.272940 0.155646  1.415552 0.096473\\
5.965334 0.209306  1.190296 0.094254\\
};
\addlegendentry{Matched fixed top-$K$}

\addplot+[
    adaptivecurve,
    /pgfplots/.cd,
    solid,
    line width=0.55pt,
    draw=AdaptiveVermilion,
    color=AdaptiveVermilion,
    mark options={
        draw=AdaptiveVermilion,
        fill=AdaptiveVermilion
    }
]
table[
    row sep=\\,
    x=overhead,
    x error=xerr,
    y=outage,
    y error=yerr
]{
overhead xerr     outage    yerr\\
1.832993 0.046676 11.647635 0.673007\\
2.172206 0.064673  8.456770 0.547113\\
2.519678 0.067474  6.290375 0.441000\\
2.952590 0.073672  4.515498 0.242623\\
3.674653 0.079191  2.525285 0.049480\\
4.014682 0.105772  2.016313 0.109189\\
4.353385 0.116184  1.755302 0.116271\\
4.883462 0.136272  1.422512 0.093996\\
5.272940 0.155646  1.259380 0.096786\\
5.965334 0.209306  0.998369 0.104710\\
};
\addlegendentry{Proposed adaptive policy}

\end{axis}
\end{tikzpicture}
}
}

\caption{Budget matched comparison of matched fixed top-$K$ selection
and the proposed adaptive policy in terms of candidate set accuracy,
average normalized received power, and outage probability.}
\label{fig:budget_matched_tradeoff_all}
\end{figure*}

\begin{figure*}[t]
\centering

\makebox[\textwidth][c]{%
\subfloat[Uncertainty-based allocation.
\label{fig:uncertainty_based_allocation}]
{%
\begin{tikzpicture}

\begin{axis}[
    name=panelA,
    width=0.72\columnwidth,
    height=0.38\columnwidth,
    scale only axis,
    xlabel={Prediction uncertainty group},
    ylabel={Mean candidate set size},
    xlabel style={font=\scriptsize},
    ylabel style={font=\scriptsize},
    tick label style={font=\tiny},
    xmin=-0.18,
    xmax=4.18,
    ymin=0.80,
    ymax=5.12,
    xtick={0,1,2,3,4},
    xticklabels={
        Lowest 20\%,
        20--40\%,
        40--60\%,
        60--80\%,
        Highest 20\%
    },
    x tick label style={
        rotate=28,
        anchor=north east,
        font=\tiny
    },
    ytick={1,2,3,4,5},
    grid=major,
    major grid style={
        draw=gray!20,
        line width=0.22pt
    },
    axis background/.style={fill=white},
    axis line style={
        draw=black!80,
        line width=0.55pt
    },
    tick style={
        draw=black!80,
        line width=0.55pt
    },
    tick align=outside,
    axis y line*=left,
    axis x line*=bottom,
    legend columns=5,
    legend style={
        font=\tiny,
        at={(0.50,1.02)},
        anchor=south,
        draw=black!35,
        line width=0.30pt,
        fill=white,
        fill opacity=1,
        text opacity=1,
        inner xsep=1.2pt,
        inner ysep=0.8pt,
        column sep=2.5pt,
        row sep=-0.2pt,
        cells={anchor=west}
    },
    legend image post style={scale=0.72},
    clip=false
]

\addplot+[
    color=TauBlue,
    line width=1.00pt,
    mark=*,
    mark size=1.40pt
]
coordinates {
    (0,1.000)
    (1,1.000)
    (2,1.035)
    (3,1.175)
    (4,1.632)
};
\addlegendentry{$\tau=.50$}

\addplot+[
    color=TauOrange,
    line width=1.00pt,
    mark=square*,
    mark size=1.35pt
]
coordinates {
    (0,1.000)
    (1,1.158)
    (2,1.246)
    (3,1.561)
    (4,1.947)
};
\addlegendentry{$\tau=.60$}

\addplot+[
    color=TauGreen,
    line width=1.00pt,
    mark=triangle*,
    mark size=1.45pt
]
coordinates {
    (0,1.000)
    (1,1.298)
    (2,1.579)
    (3,1.930)
    (4,2.193)
};
\addlegendentry{$\tau=.70$}

\addplot+[
    color=TauRed,
    line width=1.00pt,
    mark=diamond*,
    mark size=1.40pt
]
coordinates {
    (0,1.018)
    (1,1.632)
    (2,1.982)
    (3,2.070)
    (4,2.719)
};
\addlegendentry{$\tau=.80$}

\addplot+[
    color=TauPurple,
    line width=1.00pt,
    mark=triangle*,
    mark options={rotate=180},
    mark size=1.45pt
]
coordinates {
    (0,1.351)
    (1,2.035)
    (2,2.281)
    (3,2.816)
    (4,3.316)
};
\addlegendentry{$\tau=.90$}

\addplot+[
    color=TauBrown,
    densely dashed,
    line width=1.00pt,
    mark=+,
    mark size=1.60pt,
    mark options={line width=0.72pt}
]
coordinates {
    (0,1.544)
    (1,2.140)
    (2,2.491)
    (3,3.070)
    (4,3.544)
};
\addlegendentry{$\tau=.93$}

\addplot+[
    color=TauPink,
    densely dashed,
    line width=1.00pt,
    mark=x,
    mark size=1.60pt,
    mark options={line width=0.72pt}
]
coordinates {
    (0,1.754)
    (1,2.281)
    (2,2.772)
    (3,3.263)
    (4,3.807)
};
\addlegendentry{$\tau=.95$}

\addplot+[
    color=TauGray,
    densely dashed,
    line width=1.00pt,
    mark=triangle*,
    mark options={rotate=90},
    mark size=1.45pt
]
coordinates {
    (0,2.070)
    (1,2.509)
    (2,3.228)
    (3,3.649)
    (4,4.123)
};
\addlegendentry{$\tau=.97$}

\addplot+[
    color=TauOlive,
    densely dashed,
    line width=1.00pt,
    mark=triangle*,
    mark options={rotate=-90},
    mark size=1.45pt
]
coordinates {
    (0,2.281)
    (1,2.684)
    (2,3.544)
    (3,3.912)
    (4,4.368)
};
\addlegendentry{$\tau=.98$}

\addplot+[
    color=TauCyan,
    densely dashed,
    line width=1.00pt,
    mark=asterisk,
    mark size=1.60pt,
    mark options={line width=0.72pt}
]
coordinates {
    (0,2.719)
    (1,3.018)
    (2,4.053)
    (3,4.368)
    (4,4.842)
};
\addlegendentry{$\tau=.99$}

\end{axis}

\begin{axis}[
    at={(panelA.south west)},
    anchor=south west,
    width=0.72\columnwidth,
    height=0.38\columnwidth,
    scale only axis,
    xmin=-0.18,
    xmax=4.18,
    ymin=1.25,
    ymax=8.00,
    axis x line=none,
    axis y line*=right,
    axis background/.style={fill=none},
    xtick=\empty,
    ytick={2,4,6,8},
    ylabel={Group-wise equivalent overhead (\%)},
    ylabel near ticks,
    ylabel style={font=\scriptsize},
    tick label style={font=\tiny},
    axis line style={
        draw=black!80,
        line width=0.55pt
    },
    tick style={
        draw=black!80,
        line width=0.55pt
    },
    tick align=outside,
    grid=none,
    clip=false
]

\addplot[
    draw=none,
    forget plot
]
coordinates {
    (0,1.25)
    (4,8.00)
};

\end{axis}

\end{tikzpicture}
}
\hspace{0.02\textwidth}
\subfloat[Candidate count distribution.
\label{fig:candidate_count_distribution}]
{%
\begin{tikzpicture}

\begin{axis}[
    name=panelB,
    width=0.72\columnwidth,
    height=0.38\columnwidth,
    scale only axis,
    xlabel={Confidence mass threshold $\tau$},
    ylabel={Candidate set size},
    xlabel style={font=\scriptsize},
    ylabel style={font=\scriptsize},
    tick label style={font=\tiny},
    xmin=0.48,
    xmax=1.01,
    ymin=0.70,
    ymax=8.35,
    xtick={0.5,0.6,0.7,0.8,0.9,1.0},
    xticklabels={
        $0.5$,
        $0.6$,
        $0.7$,
        $0.8$,
        $0.9$,
        $1$
    },
    ytick={1,2,3,4,5,6,7,8},
    grid=major,
    major grid style={
        draw=gray!20,
        line width=0.22pt
    },
    axis background/.style={fill=white},
    axis line style={
        draw=black!80,
        line width=0.55pt
    },
    tick style={
        draw=black!80,
        line width=0.55pt
    },
    tick align=outside,
    legend columns=4,
    legend style={
        font=\tiny,
        at={(0.50,1.02)},
        anchor=south,
        draw=black!35,
        line width=0.30pt,
        fill=white,
        fill opacity=1,
        text opacity=1,
        inner xsep=1.2pt,
        inner ysep=0.8pt,
        column sep=3pt,
        cells={anchor=west}
    },
    legend image post style={scale=0.75},
    clip=false
]

\addplot+[
    color=DistGreen,
    dotted,
    line width=1.05pt,
    mark=triangle*,
    mark size=1.45pt
]
coordinates {
    (0.50,3)
    (0.60,3)
    (0.70,4)
    (0.80,4)
    (0.90,6)
    (0.93,6)
    (0.95,6)
    (0.97,7)
    (0.98,8)
    (0.99,8)
};
\addlegendentry{Max. $K$}

\addplot+[
    color=DistMagenta,
    dotted,
    line width=1.05pt,
    mark=none
]
coordinates {
    (0.48,8)
    (1.01,8)
};
\addlegendentry{$K_{\max}$}

\addplot+[
    color=DistRed,
    dotted,
    line width=1.05pt,
    mark=*,
    mark size=1.35pt,
    error bars/.cd,
    y dir=both,
    y explicit,
    error bar style={
        color=DistRed,
        line width=0.50pt
    },
    error mark=none
]
coordinates {
    (0.50,1.0) +- (0,0.000000)
    (0.60,1.0) +- (0,0.000000)
    (0.70,2.0) +- (0,0.000000)
    (0.80,2.0) +- (0,0.000000)
    (0.90,2.0) +- (0,0.000000)
    (0.93,2.6) +- (0,0.547723)
    (0.95,3.0) +- (0,0.000000)
    (0.97,3.0) +- (0,0.000000)
    (0.98,3.4) +- (0,0.547723)
    (0.99,4.0) +- (0,0.000000)
};
\addlegendentry{Median $K$}

\addplot+[
    color=DistBlue,
    dotted,
    line width=1.05pt,
    mark=square*,
    mark size=1.35pt,
    error bars/.cd,
    y dir=both,
    y explicit,
    error bar style={
        color=DistBlue,
        line width=0.50pt
    },
    error mark=none
]
coordinates {
    (0.50,2.0) +- (0,0.000000)
    (0.60,2.0) +- (0,0.000000)
    (0.70,2.0) +- (0,0.000000)
    (0.80,3.0) +- (0,0.000000)
    (0.90,3.0) +- (0,0.000000)
    (0.93,3.8) +- (0,0.447214)
    (0.95,4.0) +- (0,0.000000)
    (0.97,4.0) +- (0,0.000000)
    (0.98,5.0) +- (0,0.000000)
    (0.99,5.0) +- (0,0.000000)
};
\addlegendentry{$90$th pct. $K$}

\end{axis}

\path
    (current bounding box.south)
    ++(0,-0.42cm)
    node[inner sep=0pt] {};

\end{tikzpicture}
}
}


\caption{ {Adaptive candidate allocation under prediction uncertainty.
(a) Mean candidate set size and equivalent overhead across uncertainty groups for different $\tau$. (b) Median, $90$th-percentile, and maximum candidate set sizes versus $\tau$; error bars denote one sample standard deviation across five seeds.}}
\label{fig:combined_candidate_allocation}

\end{figure*}
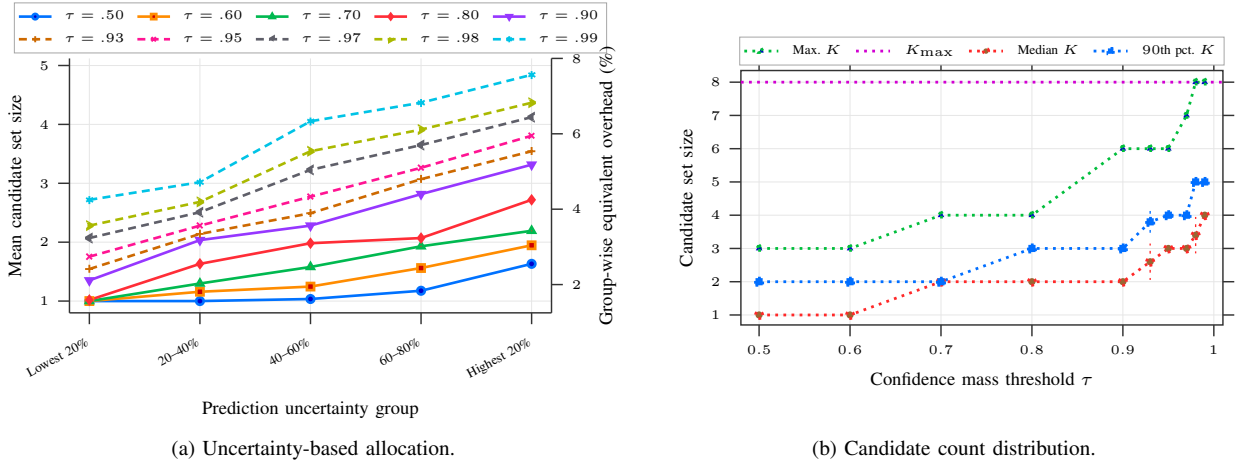

\subsection{Aggregation Across Repeated Runs}
\label{subsec:repeated_runs}

 {All experiments are repeated using $S=5$ random seeds. For a metric
$\mathcal{M}_{s}(u)$ obtained for seed $s$ at operating point $u$, where
$u$ may denote a threshold, fixed budget, or prediction step, the
reported mean and sample standard deviation are $\mu_{\mathcal{M}}(u)
=
\frac{1}{S}
\sum_{s=1}^{S}\mathcal{M}_{s}(u),$ and
\begin{equation}
\begin{aligned}
\sigma_{\mathcal{M}}(u)
&=
\sqrt{
\frac{1}{S-1}
\sum_{s=1}^{S}
\left[
\mathcal{M}_{s}(u)-\mu_{\mathcal{M}}(u)
\right]^{2}
},
\end{aligned}
\label{eq:seed_aggregation}
\end{equation}
respectively. 
Plots show $\mu_{\mathcal{M}}(u)$, with error bars representing one
sample standard deviation.}
\subsection{Calibration Metric}
\label{subsec:calibration_metric}

Calibration quality is measured using expected calibration error (ECE)
\cite{naeini2015obtaining,guo2017calibration}. For seed $s$, test sample
$i$, prediction step $r$, and temperature
$T\in\{1,T_s^{\star}\}$, define the predicted confidence and beam index
as
$q_{i,r}^{(s)}(T)
=
\max_{c\in\mathcal{C}}p_{i,r,c}^{(s)}(T)$, and
$\hat{c}_{i,r}^{(s)}(T)
=
\underset{c\in\mathcal{C}}{\arg\max}\,
p_{i,r,c}^{(s)}(T)$, respectively.
For each seed and prediction step, the confidence values are divided
into equal width bins over $[0,1]$. Let
$\mathcal{B}_{b,s,r}(T)$ contain the test samples assigned to bin $b$.
Following \cite{naeini2015obtaining, guo2017calibration}, the ECE is
\begin{equation}
\resizebox{0.98\columnwidth}{!}{$
\displaystyle
\mathrm{ECE}_{s,r}(T)
=
\sum_{b=1}^{B}
\frac{|\mathcal{B}_{b,s,r}(T)|}{N_{\mathrm{test}}}
\left|
\operatorname{acc}\!\left(\mathcal{B}_{b,s,r}(T)\right)
-
\operatorname{conf}\!\left(\mathcal{B}_{b,s,r}(T)\right)
\right|,
$}
\label{eq:ece}
\end{equation}
where $\operatorname{acc}(\cdot)$ is the empirical top $1$ accuracy, and
$\operatorname{conf}(\cdot)$ is the mean predicted confidence within a
nonempty bin; empty bins contribute zero.
The uncalibrated and calibrated values are
$\mathrm{ECE}_{s,r}(1)$ and
$\mathrm{ECE}_{s,r}(T_s^{\star})$, respectively, with lower values
indicating better calibration. ECE is computed independently for each
seed and prediction step, multiplied by $100$, and aggregated across
the five seeds using \eqref{eq:seed_aggregation}.

\section{Results and Discussion}
\label{sec:results}

This section evaluates the proposed adaptive confidence mass policy on DeepSense~6G Scenario~8 \cite{alkhateeb2021deepsense6g} using candidate set accuracy, normalized received power, outage probability, beam training overhead, prediction uncertainty, and calibration quality. The impact of probability calibration on adaptive selection is also examined. Comparisons with fixed top-$K$ selection are performed at matched average beam training overhead using \eqref{eq:budget_matching}. Results are computed over five random seeds and aggregated using \eqref{eq:seed_aggregation}, with markers and error bars denoting the mean and one sample standard deviation, respectively.

\subsubsection{Budget Matched Performance}
\label{subsec:budget_matched_results}
Figs.~\ref{fig:overhead_topk_accuracy_tradeoff}-\ref{fig:overhead_outage_tradeoff} compare the proposed adaptive policy with matched fixed top-$K$ selection under the same beam-training budgets. The adaptive policy consistently achieves higher candidate set accuracy, higher average normalized received power, and lower outage probability across all evaluated operating points. The largest candidate set accuracy gain is 3.80 percentage points at an overhead of approximately $2.52\%$, while the largest increase in average normalized received power is 0.32 percentage points at an overhead of $2.17\%$. At the same overhead, the adaptive policy also achieves its largest outage reduction of 2.06 percentage points. The performance gains are more pronounced at low beam training overhead and gradually decrease as the budget increases, since both policies approach high candidate set accuracy, recover nearly all of the optimal received power, and achieve low outage probability. The error bars further indicate that these trends are consistent across the five seeds.

\subsubsection{Adaptive Candidate Count Under Uncertainty}
\label{subsec:results_candidate_count}
Fig.~\ref{fig:combined_candidate_allocation}a quantifies how the adaptive policy allocates beams according to predictive uncertainty. Predictions are ranked by normalized predictive entropy and divided into five equal sized groups, from the lowest to the highest uncertainty.
For every evaluated value of $\tau$, the mean selected candidate count increases from the lowest to the highest uncertainty group.  {At $\tau=0.95$, for example, the mean count increases from approximately $1.75$ to $3.81$ beams. Thus, confident predictions receive smaller beam budgets, whereas uncertain predictions receive larger candidate sets. Increasing $\tau$ also raises the candidate count because more cumulative probability mass must be retained.}

Fig.~\ref{fig:combined_candidate_allocation}b summarizes the overall candidate count distribution.  {At $\tau=0.95$, the median, $90$th-percentile, and maximum counts are $3$, $4$, and $6$, respectively. The limit $K_{\max}=8$ is reached only at $\tau\in\{0.98,0.99\}$, indicating that budget saturation occurs mainly at the most restrictive confidence-mass thresholds.}

\subsubsection{Calibration Quality}
\label{subsec:calibration_quality_results}

 {For each seed, one global temperature $T_s^{\star}$ is fitted on the calibration set and applied to all prediction horizons. The fitted temperatures range from $0.877$ to $0.965$ across the five seeds. Fig.~\ref{fig:global_temperature_ece} reports the ECE defined in \eqref{eq:ece} before and after scaling. The fitted temperatures are below one for all seeds, so the calibrated distributions are sharper than the raw softmax distributions. Temperature scaling reduces ECE at every prediction horizon. Averaging the horizon values shown in the figure, ECE decreases from $5.55\%$ to $4.98\%$, an absolute reduction of $0.57\%$. The reduction indicates that global temperature scaling improves the alignment between predicted confidence and empirical accuracy.}

\begin{figure}[t]
\centering
\resizebox{0.6\columnwidth}{!}{%
\begin{tikzpicture}
\begin{axis}[
    width=0.99\columnwidth,
    height=0.61\columnwidth,
    xlabel={Prediction horizon, $r$},
    ylabel={ECE (\%)},
    xlabel style={font=\scriptsize},
    ylabel style={font=\scriptsize},
    tick label style={font=\tiny},
    xmin=0.75,
    xmax=5.35,
    ymin=0,
    ymax=9,
    xtick={1,2,3,4,5},
    ytick={0,1,2,3,4,5,6,7,8,9},
    grid=major,
    major grid style={
        draw=gray!25,
        line width=0.25pt
    },
    axis background/.style={fill=white},
    axis line style={
        draw=black!70,
        line width=0.45pt
    },
    tick style={
        draw=black!70,
        line width=0.45pt
    },
    tick align=outside,
    legend style={
        font=\tiny,
        legend columns=2,
        at={(0.03,0.965)},
        anchor=north west,
        draw=gray!60,
        fill=white,
        inner xsep=3pt,
        inner ysep=2pt,
        column sep=5pt,
        cells={anchor=west}
    },
    clip=false
]

\addplot+[
    color=VividGreen,
    densely dotted,
    line cap=round,
    line width=1.05pt,
    mark=*,
    mark size=2.1pt,
    error bars/.cd,
    y dir=both,
    y explicit,
    error bar style={
        color=VividGreen,
        line width=0.75pt
    },
    error mark=none
]
coordinates {
    (1,5.27) +- (0,1.35)
    (2,5.68) +- (0,1.43)
    (3,5.27) +- (0,1.30)
    (4,5.22) +- (0,1.45)
    (5,6.30) +- (0,2.30)
};
\addlegendentry{Uncalibrated}

\addplot[
    only marks,
    mark=-,
    mark size=4.2pt,
    mark options={
        color=VividGreen,
        line width=0.85pt
    },
    forget plot
]
coordinates {
    (1,6.62) (1,3.92)
    (2,7.11) (2,4.25)
    (3,6.57) (3,3.97)
    (4,6.67) (4,3.77)
    (5,8.60) (5,4.00)
};

\addplot+[
    color=VividRed,
    densely dotted,
    line cap=round,
    line width=1.05pt,
    mark=square*,
    mark size=2pt,
    error bars/.cd,
    y dir=both,
    y explicit,
    error bar style={
        color=VividRed,
        line width=0.75pt
    },
    error mark=none
]
coordinates {
    (1,4.47) +- (0,0.62)
    (2,5.38) +- (0,1.25)
    (3,4.48) +- (0,0.68)
    (4,4.76) +- (0,0.90)
    (5,5.80) +- (0,1.25)
};
\addlegendentry{Global temperature}

\addplot[
    only marks,
    mark=-,
    mark size=4.2pt,
    mark options={
        color=VividRed,
        line width=0.85pt
    },
    forget plot
]
coordinates {
    (1,5.09) (1,3.85)
    (2,6.63) (2,4.13)
    (3,5.16) (3,3.80)
    (4,5.66) (4,3.86)
    (5,7.05) (5,4.55)
};

\node[
    anchor=south west,
    fill=white,
    fill opacity=0.98,
    text opacity=1,
    draw=black!35,
    line width=0.35pt,
    rounded corners=1pt,
    inner xsep=2.5pt,
    inner ysep=2pt,
    font=\fontsize{4.1}{4.8}\selectfont
]
at (rel axis cs:0.075,0.055)
{%
\begingroup
\setlength{\tabcolsep}{2.3pt}
\renewcommand{\arraystretch}{1.05}
\begin{tabular}{c c c}
\textbf{Seed} &
\textbf{Raw softmax $T$} &
\textbf{Fitted $T_s^{\star}$}\\
\hline
1 & 1.000000 & 0.965418\\
2 & 1.000000 & 0.955028\\
3 & 1.000000 & 0.889806\\
4 & 1.000000 & 0.891226\\
5 & 1.000000 & 0.877329
\end{tabular}
\endgroup
};

\end{axis}
\end{tikzpicture}
}
\caption{ {ECE before and after global temperature scaling. Markers show
the mean across five seeds, and error bars show one sample standard
deviation.}}
\label{fig:global_temperature_ece}
\end{figure}
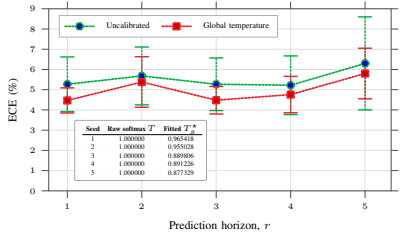

\subsubsection{Effect of Calibration on Adaptive Selection}
\label{subsec:calibration_effect}

 {Temperature scaling preserves beam ordering and thus leaves fixed top-$K$ candidate sets unchanged. In contrast, the adaptive policy depends on cumulative probability mass, so calibration can alter the selected set size even when the ranking remains unchanged.}

 {Fig.~\ref{fig:temperature_overhead} shows that probability calibration reduces the mean beam-training overhead across all evaluated values of $\tau$. As a representative example, at $\tau=0.95$, the mean overhead decreases from $4.65\%$ to $4.35\%$, corresponding to a reduction in the average candidate set size from $2.98$ to $2.79$ beams. This represents
an absolute overhead reduction of approximately $0.30\%$. A similar trend is observed for the remaining confidence mass thresholds, indicating that the benefit
is not specific to $\tau=0.95$. Furthermore,
Fig.~\ref{fig:seedwise_temperature_overhead} shows that the overhead reduction is consistently observed across all evaluated random seeds.}

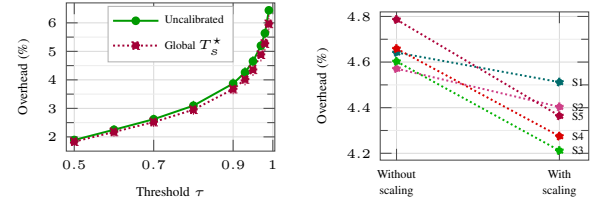
\begin{figure}[t]
\centering

\makebox[\columnwidth][c]{%
\subfloat[Overhead across $\tau$.
\label{fig:temperature_overhead}]
{%
\begin{tikzpicture}
\begin{axis}[
    width=0.49\columnwidth,
    height=0.40\columnwidth,
    xlabel={Threshold $\tau$},
    ylabel={Overhead (\%)},
    xlabel style={font=\tiny},
    ylabel style={font=\tiny},
    tick label style={font=\tiny},
    xmin=0.48,
    xmax=1.005,
    ymin=1.5,
    ymax=6.7,
    xtick={0.5,0.7,0.9,1.0},
    ytick={2,3,4,5,6},
    minor x tick num=1,
    minor y tick num=1,
    grid=both,
    major grid style={
        draw=gray!30,
        line width=0.22pt
    },
    minor grid style={
        draw=gray!15,
        line width=0.15pt
    },
    axis line style={
        draw=black!70,
        line width=0.40pt
    },
    tick style={
        draw=black!70,
        line width=0.40pt
    },
    legend style={
        font=\fontsize{4.3}{4.8}\selectfont,
        at={(0.5,0.98)},
        anchor=north,
        draw=black!30,
        fill=white,
        inner xsep=1.5pt,
        inner ysep=1.2pt,
        row sep=0.2pt,
        cells={anchor=west}
    },
    clip=false
]

\addplot[
    green!60!black,
    solid,
    line width=0.75pt,
    mark=*,
    mark size=1.25pt,
    error bars/.cd,
    y dir=both,
    y explicit,
    error bar style={line width=0.35pt},
    error mark options={
        line width=0.35pt,
        mark size=1.1pt
    }
]
coordinates {
    (0.50,1.890395) +- (0,0.029555)
    (0.60,2.250102) +- (0,0.054434)
    (0.70,2.619291) +- (0,0.054483)
    (0.80,3.094617) +- (0,0.069058)
    (0.90,3.869800) +- (0,0.078783)
    (0.93,4.261725) +- (0,0.082588)
    (0.95,4.652325) +- (0,0.082370)
    (0.97,5.197695) +- (0,0.062344)
    (0.98,5.634380) +- (0,0.047200)
    (0.99,6.439947) +- (0,0.015852)
};
\addlegendentry{Uncalibrated}

\addplot[
    purple!85!black,
    densely dotted,
    line width=0.75pt,
    mark=square*,
    mark size=1.25pt,
    error bars/.cd,
    y dir=both,
    y explicit,
    error bar style={line width=0.35pt},
    error mark options={
        line width=0.35pt,
        mark size=1.1pt
    }
]
coordinates {
    (0.50,1.832994) +- (0,0.046677)
    (0.60,2.172206) +- (0,0.064673)
    (0.70,2.519678) +- (0,0.067473)
    (0.80,2.952589) +- (0,0.073672)
    (0.90,3.674653) +- (0,0.079191)
    (0.93,4.014681) +- (0,0.105772)
    (0.95,4.353384) +- (0,0.116184)
    (0.97,4.883463) +- (0,0.136272)
    (0.98,5.272941) +- (0,0.155645)
    (0.99,5.965334) +- (0,0.209306)
};
\addlegendentry{Global $T_s^{\star}$}

\end{axis}
\end{tikzpicture}
}
\hspace{0.005\columnwidth}
\subfloat[Seed results at $\tau=0.95$.
\label{fig:seedwise_temperature_overhead}]
{%
\begin{tikzpicture}
\begin{axis}[
    width=0.49\columnwidth,
    height=0.40\columnwidth,
    ylabel={Overhead (\%)},
    ylabel style={font=\tiny},
    tick label style={font=\fontsize{4.5}{5}\selectfont},
    xmin=-0.10,
    xmax=1.18,
    ymin=4.17,
    ymax=4.82,
    xtick={0,1},
    xticklabels={
        \shortstack{Without\\scaling},
        \shortstack{With\\scaling}
    },
    ytick={4.2,4.4,4.6,4.8},
    minor y tick num=1,
    grid=both,
    major grid style={
        draw=gray!30,
        line width=0.22pt
    },
    minor grid style={
        draw=gray!15,
        line width=0.15pt
    },
    axis line style={
        draw=black!70,
        line width=0.40pt
    },
    tick style={
        draw=black!70,
        line width=0.40pt
    },
    clip=false
]

\addplot[
    teal!85!black,
    densely dotted,
    line width=0.75pt,
    mark=*,
    mark size=1.3pt
]
coordinates {
    (0,4.642128)
    (1,4.512133)
};
\node[
    anchor=west,
    font=\fontsize{4.2}{4.6}\selectfont
]
at (axis cs:1.02,4.512133) {S1};

\addplot[
    magenta!85!black,
    densely dotted,
    line width=0.75pt,
    mark=*,
    mark size=1.3pt
]
coordinates {
    (0,4.571269)
    (1,4.403548)
};
\node[
    anchor=west,
    font=\fontsize{4.2}{4.6}\selectfont
]
at (axis cs:1.02,4.403548) {S2};

\addplot[
    green!70!black,
    densely dotted,
    line width=0.75pt,
    mark=*,
    mark size=1.3pt
]
coordinates {
    (0,4.602875)
    (1,4.211867)
};
\node[
    anchor=west,
    font=\fontsize{4.2}{4.6}\selectfont
]
at (axis cs:1.02,4.211867) {S3};

\addplot[
    red!85!black,
    densely dotted,
    line width=0.75pt,
    mark=*,
    mark size=1.3pt
]
coordinates {
    (0,4.658952)
    (1,4.275081)
};
\node[
    anchor=west,
    font=\fontsize{4.2}{4.6}\selectfont
]
at (axis cs:1.02,4.275081) {S4};

\addplot[
    purple!90!black,
    densely dotted,
    line width=0.75pt,
    mark=*,
    mark size=1.3pt
]
coordinates {
    (0,4.786398)
    (1,4.364294)
};
\node[
    anchor=west,
    font=\fontsize{4.2}{4.6}\selectfont
]
at (axis cs:1.02,4.364294) {S5};

\end{axis}
\end{tikzpicture}
}
}

\caption{ {Effect of global temperature scaling on adaptive beam training
overhead.}}
\label{fig:calibration_effect_overhead}
\end{figure}

\section{Conclusion}
This letter examined whether calibrated prediction confidence can support adaptive beam-budget allocation in vision-aided beam tracking. The results show that the proposed policy provides a more favorable balance between communication reliability and beam-training overhead than budget-matched fixed top-$K$ selection. Its candidate count increases with prediction uncertainty, while the imposed limit $K_{\max}$ keeps the maximum overhead controlled. The calibration study also shows that temperature scaling improves the quality of the predicted probabilities and reduces the beam budget required by the adaptive rule, even though the beam ranking remains unchanged. Since the underlying predictor is neither modified nor retrained, the method can be introduced as a lightweight decision layer for existing beam prediction architectures. Future work may investigate conformal prediction-set calibration to associate the confidence threshold with an explicit coverage guarantee and evaluate the approach across additional propagation scenarios and beam codebooks \cite{romano2020classification}.


\endgroup

\bibliographystyle{ieeetr}
\bibliography{ref.bib}

\end{document}